# On Security Measures for Containerized Applications Imaged with Docker

Samuel P. Mullinix, Erikton Konomi, Renee Davis Townsend, Reza M. Parizi
College of Computing and Software Engineering, Kennesaw State University, GA, 30060 USA
{smullini, ekonomi, ldavisto}@students.kennesaw.edu, rparizi1@kennesaw.edu

*Abstract*—Linux containers have risen in popularity in the last few years, making their way to commercial IT service offerings (such as PaaS), application deployments, and Continuous Delivery/Integration pipelines within various development teams. Along with the wide adoption of Docker, security vulnerabilities and concerns have also surfaced. In this survey, we examine the state of security for the most popular container system at the moment: Docker. We will also look into its origins stemming from the Linux technologies built into the OS itself; examine intrinsic vulnerabilities, such as the Docker Image implementation; and provide an analysis of current tools and modern methodologies used in the field to evaluate and enhance its security. For each section, we pinpoint metrics of interest, as they have been revealed by researchers and experts in the domain and summarize their findings to paint a holistic picture of the efforts behind those findings. Lastly, we look at tools utilized in the industry to streamline Docker security scanning and analytics which provide built-in aggregation of key metrics.

*Keywords—docker; container; virtualization; security; metrics*

## I. Introduction

Docker containers have become an integral part of modern information technology (IT) infrastructure. They are used in combination with, or as a replacement of, traditional Virtual Machines (VM), offering a service or application that is lighter on resources yet still isolated from other services on the same system (or VM) The rapid expansion of this new technology also leads to an entirely new and different attack surface compared to application infrastructure and architectures used just a decade ago. Tripwire, a leading software vendor in the IT security and compliance area, performed a "State of Container Security Report" in 2019 that focused on surveying over 300 IT professionals working at companies with over 100 employees. These 300 professionals were directly responsible for maintaining environments with containers and it was reported that 94% of the respondents claimed to be concerned about container security, and 60% had a container security incident in the last year [1]. Additionally, managing the unique security situations created by Docker is important not only because security professionals are concerned about application security, but it is also crucial to understand as Docker powers some of the world's largest companies such as AT&T, Netflix, Adobe, and PayPal [2].

In this paper, we focus specifically on the security concerns surrounding Docker and aim to provide the reader with an overview of the current state of affairs in this domain. We explore research from various academic bodies and technical conferences around the world, leveraging their findings to provide a complete picture of the most important security aspects and metrics.

First, we provide the necessary background for understanding how Docker emerged from LXC (Linux Container), and how both rely on the Linux kernel. Then we systematically analyze security concerns, Docker images, and security prevention techniques via static and dynamic analysis methods. Finally, industry-leading tools and automation services that help with constant monitoring and guarding against exploitation attempts are introduced.

Based on these methods of Docker security analysis, we have determined that Dockers relatively new yet widespread adoption leaves opportunity for security evaluation. Metrics derived from CPU, network, and memory statics provide a great start for anomaly detection techniques. However, Dockers unique infrastructure, including how it sits on top of a host operating system, means that there are some unique security implications that require custom tooling and new methods of analysis to ensure that results are actionable and monitorable.

## II. Docker and architectural security

### A. Linux Security Overview

At its core, a Docker image is a tightly packaged GNU/Linux kernel, with all the necessary libraries and utilities needed to run the application in question (e.g. MySQL database instance). In such, it is important to examine the state of Linux security and understand the technologies that safeguard the host operating system from malicious attacks before diving into the Docker specifics.

The basic, although outdated, security features of the Linux kernel are inherited from Unix itself, under the name of *DAC* or *Discretionary Access Control*. Simply put, this is the privilege-based system that allows the existence of multiple users and groups, allowing each access to files, system locations, and applications according to this ownership scheme. A system administrator (or *superuser*) can override these privileges however, posing a security risk.

One of the main extensions to this basic schema is the usage of a *cryptography* API (Application Programming Interface) via a kernel module to allow further hardening of privileges throughout the system installation. Another enhancement is the usage of *LSM* or *Linux Security Modules* which further builds upon *DAC* and introduces an intermediate layer for applications to register through and receive callbacks from. This ensures a decoupled system is maintained between the user privileges and the application privileges at runtime via the implementation of isolation rules and policies.

Two prominent and widely-used *LSM* implementations are *SELinux* and *AppArmor*. These are found in popular Linux distributions such as Fedora and Ubuntu Linux, commonly used throughout the IT infrastructure domain. Both of them have evolved over time to accommodate the ever-changing landscape of software applications while mitigating security risks and maintaining a strong capability with respect to virtualization environments, one of which is Docker.

The main drawback of both these tools is their reliance on user-defined rules and policies. As such, the usability of the tool becomes immediately critical to the configuration of the system. Schreuders, McGill and Payne [3] have conducted an empirical study on this subject, introducing a graphical toolkit to aid administrators in the exploration and implementation of current SELinux policies using visualization. The study, although limited in scope, proved that visual aids enhanced the usability of SELinux as a whole and helped administrators identify policy violations.

A similar study conducted by the same group of researchers the following year utilized a detailed comparison of *SELinux*, *AppArmor* and a third LSM implementation by the name of *FBAC-LSM* [4]. This comparison found that out of the three, *FBAC-LSM* was favored amongst participants due to the friendly user interface that by extension allowed better policy implementation and inspection. However, this security framework is no longer being utilized.

## B. Linux Features that Enable Docker

Going further into the security measures implemented into the Linux kernel, there are two specific features that directly enable the existence of containers, such as Docker and LXC; these are: *namespaces* and *control groups*.

*Namespaces* are used to isolate processes on several levels and provide the necessary mechanisms to orchestrate complex infrastructure. This is not restricted to containers and virtualization alone but can be used for other applications as well. There are currently six such isolation levels, or namespaces implemented:

1. mnt (mount point, filesystem)
2. pid (processes)
3. net (network stack)
4. ipc (system interrupt calls)
5. uts (hostname)
6. user (UIDs)

By allocating a dedicated namespace (or several) to a process, a certain level of isolation, or sandboxing, is guaranteed. A container is essentially a process running on the host OS, and multiple containers can run in isolation using this method.

While *namespaces* ensure isolation, *control groups* (usually called just *cgroups*) are responsible for resource management and provide the necessary mechanisms to programmatically allocate resources to a process. Through the use of a virtual file system (VFS), each *cgroup* can be assigned a limited number of resources by the administrator to (a) ensure they do not exceed available host resources and (b) not starve system resources for other applications. The resources are usually CPU, memory or network related, but there are hardware-oriented policies as well, such as a device control group that can restrict read/write access to physical devices.

## C. Docker Security

Despite the facilities mentioned in the previous section, there are more implementation-specific details to be taken into consideration when evaluating the state of container security.

One comprehensive study by Dua, Raja and Kakadia evaluated various container technologies from the scope of suitability in a PaaS (Product as a Service) environment, while also comparing them to existing virtualization technologies, such as Xen and KVM [5]. A high-level comparison is provided in Table 1. When compared with traditional virtualization options, containers have several advantages, mainly in terms of performance and overall manageability. On the security front, all types of containers can benefit from *Mandatory Access Control* on the host OS via tools such as *SELinux* and *AppArmor* as mentioned in the previous section.

*Table 1. VM and Container Feature Comparison from [3]*

| Parameter | Virtual Machines | Containers |
|---|---|---|
| Guest OS | Each VM runs on virtual hardware and Kernel is loaded into its own memory region | All the guests share the same OS and Kernel. The kernel image is loaded into the physical memory |
| Communication | Occurs through Ethernet Devices | Standard IPC mechanisms like signals, pipes, and sockets. |
| Security | Depends on the implementation of Hypervisor | Mandatory access control can be leveraged |
| Performance | Virtual Machines suffer from a small overhead since the machine instructions and translated from the Guest to the Host OS | Containers provide near-native performance compared to the underlying Host OS |
| Isolation | Sharing libraries and files between guests and guest hosts are not possible | Subdirectories can be transparently mounted and can be shared |
| Startup time | VM's take several minutes to boot up | Containers can be started in several seconds since the host OS is already running |
| Storage | Vm's take more storage since the OS kernel and its associated programs have to be installed, stored, and executed | Containers take a lower amount of storage since the base OS is shared |

Docker specifically builds upon *LXC* (Linux Containers), a lightweight kernel container implementation. Additionally, it leverages the same aforementioned Linux mechanisms, *namespaces* and *cgroups*, while also aiming to resolve two more issues: (a) mapping the root user of the container to the non-root user of Docker instance, and (b) allowing the Docker daemon to run as a non-root user. Table 2 illustrates the differences between Docker and LXC, as well as two other popular container technologies.

*Table 2. Container Implementation Comparison from [5]*

| Parameter | LXC | Warden | Docker | OpenVZ |
|---|---|---|---|---|
| Process Isolation | Uses pid namespaces | Uses pid namespaces | Uses pid namespaces | Uses pid namespaces |
| Resource Isolation | Uses cgroups | Uses cgroups | Uses cgroups | Uses cgroups |
| Network Isolation | Uses net namespaces | Uses net namespaces | Uses net namespaces | Uses net namespaces |
| Filesystem Isolation | Uses chroot | Uses Overlay file system with overlayfs | Uses chroot | Uses chroot |
| Container Lifecycle | Tools, lcx-create, lcx-stop, lcx start, to create start and stop containers | Containers are managed but running commands on a warden client and warden server | Uses Docker daemon and a client to manage containers | Uses vzctl to manage container lifecycle |

In fact, resource isolation mechanisms are the same as discussed in the previous section (*cgroups*), and the various containers only differ in the Filesystem Isolation and Container Lifecycle categories. These are implementation specific details that provide the abstraction layers needed for each container and are depicted in Figure 1. This graph shows the reliance of Docker to LXC and the container specific changes including the UnionFS and Image portions. The topic of a Docker image is of particular interest from a security perspective and is discussed in the next section.

### III. DOCKER IMAGE SECURITY & METRICS

One important aspect of how Docker images are composed involves the regulation and creation of Docker Images. Essentially, these image files are generated based on a set of instructions provided to the Docker Engine. Together, the creation of an image and all of the aspects in which it runs contributes to interesting metrics that can also involve malicious or outdated images. Additionally, common frameworks such as CVE (Common Vulnerability Assessment) and tools like Docker Notary can be used to find these vulnerabilities.

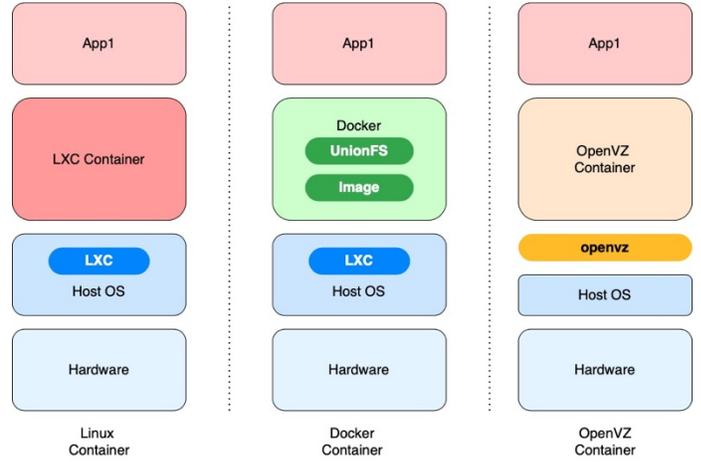

*Figure 1. Comparing Various Containers from [5]*

Image files contain the needed code, configuration mapping, and packages that are needed to run an application [6]. Therefore, and based on any of these components, an image can be malicious or even house vulnerable packages that could have security implications. Currently, there are many different ways and tools to discover these items, but one topic of interest is static and dynamic analysis.

#### A. Malicious/Vulnerable Images

One important aspect of using images is where they are hosted. Similar to how code is hosted within GitHub, many developers register their images for reuse within Docker Hub. For example, if a .NET application is built with an image, then it is possible that another Docker Hub user may encounter the same scenario and want to reuse the image, only changing or adding the specific configurations/code needed for his or her application. This new method encourages reusability and file sharing [6]. However, this means that these images are developed independently and then trafficked out by a litany of organizations. Therefore, the image file can turn into a dependency that raises concerns [6]. Acknowledging that this is a problem, even architects at Docker are encouraging developers and image composers to assume that all of these distributed pipelines are "actively malicious" [6][7].

To combat this issue, Docker has introduced a utility known as *Notary* which is used to safely publish key content over a distributed network [7]. To do this, it relies on the following metrics in the left column of Table 3 in comparison to those from Brady [6] on the right.

*Table 3. Comparison of Brady et. al and Winkle Analysis services from [6][8]*

| Winkle Analysis [8] | Brady et. al Analysis [6] |
|---|---|
| Survivable Key Compromise | CVE Analysis |
| Freshness Guarantee | Dynamic Analysis |
| Configurable Trust Thresholds | Static Analysis |
| Single delegation | Port Scans |
| Existing Distribution | Process Monitoring |
| Untrusted mirrors and transport | |

This framework allows for the creation of a content trust network that can be enabled in a container registry. With integration to a Docker Trust Server, one can use the Notary tool to pull, push, and sign images [6]. What this ensures is that if there is any tampering or modifications to the trusted image, then the Notary service will be able to report the malicious item during the scan and verification.

Malicious images can be detected using vulnerability assessment and virus tools [18]. Additionally, these malicious images frequently use bash scripts to establish a Secure Shell Tunnel (SSH) to report back to a central network [6]. To prevent this, a Docker in Docker analysis can be used to capture metrics relevant to this such as networking, CPU, and memory statistics. This approach is known as dynamic analysis.

### B. Dynamic Analysis & CVE

In comparison to using tools built into Docker, dynamic analysis observes how the container and image behave while the instance is executing [6]. While dynamic analysis takes more time than static analysis, it can provide better and more actionable results [6].

In the paper "Docker Container Security in Cloud Computing," Brandy et. al achieved a static and dynamic analysis pipeline leveraged against Docker images by combining an API that uses the tool CoreOS Clair and VirusTotal to scan against CVE (Common Vulnerabilities and Exposures) [6]. See Figure 3 that shows how the image is loaded into the API and uses the two tools to output results.

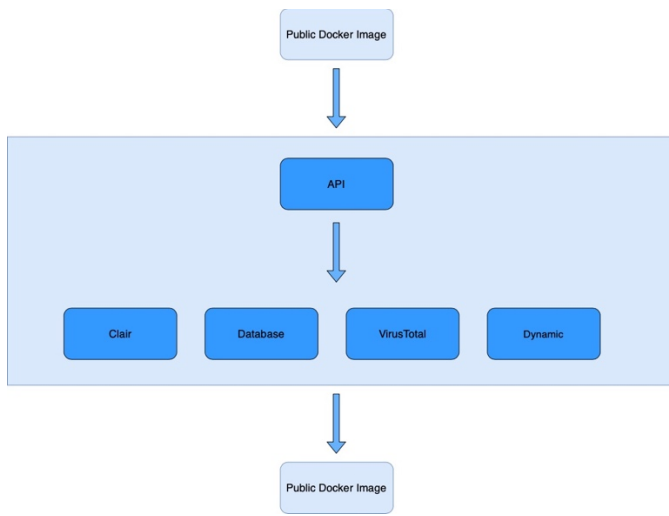

*Figure 3. Dynamic & Static Analysis Scan Using Clair and VirusTotal from Brandy et. al [6].*

While Clair provides a static analysis (more on this in the next section) screening, virus scanning tools such as VirusTotal perform static analysis via signature checks and they also run dynamic analysis within emulators [6]. This method relies on three different metrics:

1. Network analysis
    a. Packet capture
    b. Connection time alive amount
    c. Open ports
2. File system changes
    a. Changes to image
    b. New binaries
3. CPU statistics
    a. RAM consumption
    b. CPU usage

One additional type of dynamic analysis mentioned by Brandy et al. [6] is the use of Shodan to scan inbound/outbound IP ranges to determine whether the connection originates from an SSH or TOR daemon. Additionally, this tool can report open ports and services running. In summary, dynamic analysis requires more resources and evaluation since it runs during the program execution compared to doing static analysis before publishing an image. However, it is excellent for providing actionable results based on relevant image metrics.

### C. Static Analysis

Static analysis of a Docker image involves checking the image while it is at rest to ensure that vulnerabilities or security issues are detected before the image is running. Figure 4 shows an example of this from Brandy et. al where image analysis is performed at rest before pushing to a Docker Registry.

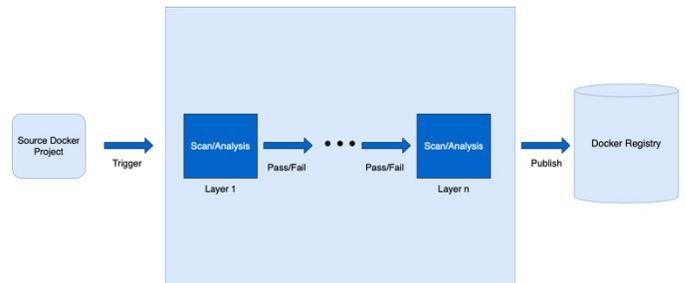

*Figure 4. Static Analysis Workflow for Docker Security Metrics from Brandy et. al [6].*

CVE is essentially a framework and data log that assists in this static analysis workflow. It provides an extensive catalog of key known vulnerabilities. Included in this report is a list of comprehensive analysis of known vulnerabilities along with their severity scores and impact information [7]. This is crucial to Docker because it creates an actionable reference point for static analysis. Additionally, it was found that if a vulnerability within a Docker image was detected, then it was more likely that action would be taken to address it [9].

*Table 4. Clair CVE Metric Categories from [6][7]*

| Category | Severity Level |
|---|---|
| Unknown | 0 |
| Negligible | 1 |
| Low | 2 |
| Medium | 3 |
| High | 4 |
| Critical | 5 |
| Defcon1 | 6 |

Brandy et al. utilized CoreOS Clair and VirusTotal within the first and second layers of their static analysis scan to extract the needed results. Clair is able to categorize the known CVE's into seven different metrics as shown in Table 4.

Brandy et al. [6] then use static analysis in their CI/CD pipelines to determine how malicious an image is by aggregating two values: a threshold based on the above severity level, and a vulnerability count. For an image to pass the metric scan, an image will be approved if and only if there were less than 50 vulnerabilities of severity 4 or category "High" [6]. In comparison, this type of CVE analysis aligns with works completed by Zerouali et al. [10] who found that "users who are more concerned by image security focused on scanning for simple Common Vulnerabilities and Exposures (CVE)." Finally, the following list identifies the metrics that Brandy et al. identified in relation to static analysis [6]:

- Code issues
    - Number of source code bugs
    - Lines of unreachable code
    - Number of undefined variables
    - Amount of variable misuse
    - Numbed of uncalled functions
    - Amount of improper memory usage
    - Number of out of bounds exceptions
- File issues
    - Signature based file names
    - Hashes on files
    - Malicious file types

The combination of these metrics provides the basis for observing and determining the vulnerabilities contained within an image using static analysis.

### D. Outdated Images

In comparison to static analysis, Zerouali et al. [10] reviewed and performed a combination of static and dynamic analysis for outdated packages in the paper "On the Relation between Outdated Docker Containers, Severity, Vulnerabilities and Bugs." Zeriouali and the authors describe that Docker Hub hosts over 1.5 million images in its repositories. These repositories usually house both open-sourced and community certified images from regular vendors [10]. However, instead of looking for images that have been maliciously impacted, Zerouali et al. look into how an old version of an image may have outdated packages that contain security vulnerabilities and bugs that impact the image [10]. Even though the code may be updated, if the image of the Docker container is not, then it is possible that a production image would have bugs and flaws that could compromise security.

Zerouali et al. also focused on Common Vulnerabilities and Exposures (CVE) ratings and found that in a similar fashion to Brandy et al., users who are concerned about vulnerabilities apply efforts to comparing vulnerabilities on the CVE list [10][6]. This backs up some of the supporting claims from Brandy et al., but fundamentally Zerouali et al. focused on different research questions. See Table 5 for the full list of research questions posed by Zerouali et al.

Table 5. Zerouali et al. Research Questions from [10]

| Question Number | Question |
|---|---|
| RQ0 | "How often are Docker images updated?" |
| RQ1 | "How outdated are container packages?" |
| RQ2 | "How vulnerable are container packages?" |
| RQ3 | "To which extent do containers suffer from bus in packages?" |
| RQ4 | "How long do bugs and security vulnerabilities remain unfixed?" |

Zerouali et al. are able to extract the needed information to answer the above research questions using the below method in Figure 5.

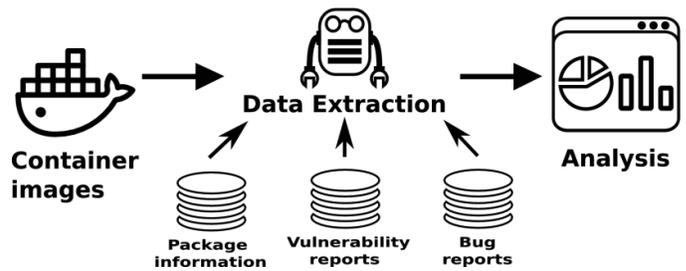

Figure 5. Zerouali et al. Outdated Image Analysis Diagram [10]

This method relies on scanning Docker images to assess the package, vulnerability, and bug reports for further analysis. Specifically, the tool "Debian Security Bug Tracker" is used. This tool is a managed bug defect list that combines CVE data and can be cross correlated with images pushed to the scanned container [6]. Additionally, Zerouali et al. utilized the following outputs from the Debian report to summarize their analysis [10]:

- Affected source packages
- Vulnerability Severity
- Vulnerability Status
- Package Version
- Debian Bug ID
- Affected Debian Distribution

Zerouali et al. also found that "more than half of the scanned Docker images had not been updated for four months" [10]. In regard to RQ1, it was found that one out of every five packages installed from a Docker image is outdated [10]. Zerouali et al. also determined the proportions of vulnerabilities within image packages. It was determined that while medium and high vulnerabilities had the highest percent of resolved cases, all levels of vulnerability contained a percentage of open issues [10].

From this finding, it was hypothesized that images or containers with a higher number of outdated packages would, therefore, have a higher amount of vulnerabilities. Zerouali et al. found a positive correlation between the number of outdated packages and the total located vulnerabilities across multiple versions of Debian. Additionally, older versions correlated with a higher vulnerability count [10].

With these findings, Zerouali et al. were able to conclude that nearly half of all vulnerabilities are not being fixed, and furthermore, many containers have high severity issues [10]. In summary, it provides insight into known Docker security pattern analysis by laying a groundwork to bridge CVE correlation with outdated images and metrics.

## IV. DOCKER THREATS AND ATTACKS

Docker contains different types of potential attacks that can cause critical data dissemination occurring primarily between Docker to Docker or Container to Container communications. This section will cover the threat model metrics and channel attacks as proposed by Yang Luo et al. [11] along with a graph-based anomaly detection system as described by Chengzhi Lu et al. [12].

### A. Threat Model

Luo et al. identified the container threat models that are officially documented paths for intra-container communication. These paths are important because they define key metrics that directly correlate to the security of a Docker container and the functionality it encapsulates and permits [11]. Refer to Table 6 for the three different types of documented container-container communication pathways:

*Table 6. Container to Container Official Communication Paths from [11]*

| Collocated Container Communication Type | Area of Application |
|---|---|
| Storage Path Mapping | Supplied with collocated container storage mounting. |
| Port Mapping | Network intra-communications |
| Layer-2 Connection | Virtual intra-container network bridge |

Storage path mapping communication allows for additional arguments to be passed to the container. When allowed, this file path is mapped as storage to be supplied to the host operating system and it is then mapped between the two containers [11]. Port Mapping allows for a mapping of a certain port or ports from one container to another through an underlying TCP or UDP connection [11]. Finally, Luo et al. identified that a Layer-2 connection is possible if a Layer-2 network connection via a virtual bridge is enabled or established.

Knowing this, Luo et al. proposed that if these communication types are not properly understood, then it would be possible to retrieve secrets or protected information from one container through a collocated container. Figure 6 shows this process where a malicious resource through the internet could access Container B, go through the Docker and OS Channel through the previously mentioned communication models, and then retrieve access to Container A's secret information.

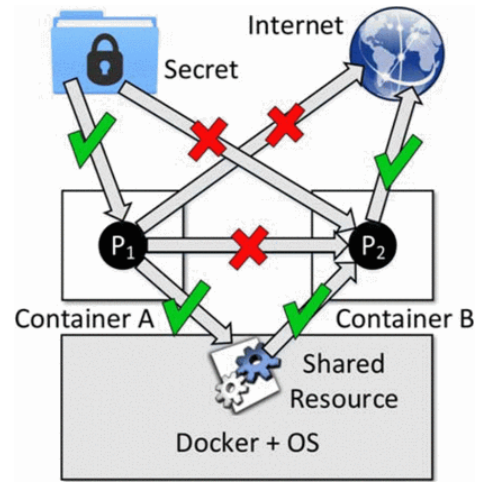

*Figure 6. Threat model utilizing channel attack between Docker containers from [11]*

This information is crucial because there is an opportunity to observe the potential types of threats presented in the model Luo et al. established by investigating the metrics in these channels of attack.

### B. Channel Attacks

Since containers all run on the same machine, they obviously share a significant amount of hardware resources like CPU utilization and memory. With this in mind, and since containers can view the changes in memory usage per process and container, Luo et al. identified this as one crucial metric to analyze known as "hardware occupancy channels" [11]. Another option that relates to file system mapping is to identify the disk operations on the mounted hosts shared between the two containers. This structure is mounted on both containers and shared as a filesystem from the physical host [11]. Because of this relation, Luo et al. proposed that the used size of the shared disk can be observed to monitor for process changes in "non-privileged containers."

Another type of channel attack mentioned by Luo et al. is the KMB (Kernel Message Buffer) Channel and Information Leak attack. This attack involves utilizing and gaining access to kernel-level messages that are sensitive to the machine. With this type of attack, Luo et al. identified the following metrics can be gained maliciously [11]:

- CPU details
- Linux Kernel Version
- BIOS information
- Hypervisor version
- Memory information
- Systemd daemon running status
- Security modules
- Bluetooth and network adapter details

In order to prevent these attacks, Luo et al. identify that these attack services can be resolved if the "full" privilege mode parameter is not utilized. This will decrease the attack surface of the container by reducing the number of potential attack channels from 37 to 14 [11]. Luo et al. identified that these remaining 14 capabilities are granted by default and they are listed below in Table 7.

*Table 7. Container to Container Official Communication Paths from [11]*

| Group | Isolated? | Capabilities |
|---|---|---|
| System (1/9) | No | CAP_AUDIT_WRITE |
| File Related (7/11) | Yes | CAP_CHOWN, CAP_DAC_OVERRIDE, CAP_FOWNER, CAP_FSETID, CAP_MKNOD, CAP_SYS_CHROOT, CAP_SETFCAP |
| Process Related (4/6) | Yes | CAP_KILL, CAP_SETGID, CAP_SETUID, CAP_SETCAP |
| Network Related (2/3) | Yes | CAP_NET_BIND_SERVICE |

In conclusion, Luo et al. found that the management of these capabilities along with monitoring the KMB and utilizing a security policy are key factors in adhering to a reduction application attack surface [11].

*C. Graph Based Anomaly Detection*

In comparison to the analysis by Luo et al., Lu et al. have determined a "novel graph-based anomaly detection" system that can avoid using traditional detection methods to locate system vulnerabilities and issues. Lu et al. mention that most other Docker security scanning tools or methods rely on the following metrics to determine if a container is acting in an anomalous state:

- CPU usage
- Memory usage
- Network input/output rates

However, and due to the common nature of n-tiered applications, these metrics can be unreliable and cause false positives [12]. This is because such systems may behave in dynamic manners and this can lead to results or readings that look like security anomalies, when in fact they represent normal system behavior (e.g. spikes in CPU usage based on business strategies). In comparison to the analysis by Luo et al and others, Lu et al. utilize an edge weight system state graph that represents the multiple tiers of the application. This allows for multiple components of a system to be represented based on functionality and average system utilization or role.

Furthermore, the system relies mostly on anticipated component response time between the different nodes of the system. See Figure 7 for the request time analysis of a normal three-tiered application.

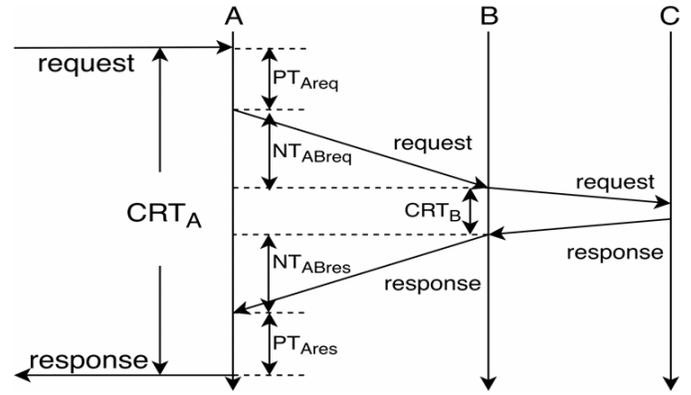

*Figure 7. Threat model utilizing channel attack between Docker containers from [12]*

Here the total response time is defined as [12]:

$$CRT_A = CRT_B + NT_{ABreq} + NT_{ABres} + PT_{Areq} + PT_{Ares}$$

Where these values coordinate with the following values in Table 8.

*Table 8. Three-tier application Response Time Definitions from [12]*

| Value | Definition |
|---|---|
| CRTA | Total response time for component A |
| CRTB | Total response time for component B |
| NTABreq | Request time between component A and component B |
| NTABres | Response time between component A and component B |
| PTAreq | Component A request processing time |
| PTAres | Component A processing response time |

From here, it can be extrapolated that each component will have a weight difference based on the calculated Euclidian distances between the edges. Based on these weight differences, the response time and graph similarity can be compared over time and is represented in a stable system. Lu determined that when plotted, the graph similarity of these components across a sample point stays approximately straight.

Additionally, Lu et al. tested the system with a CPU, memory, and network anomaly. The result was that the response time of the nodes in all three scenarios demonstrated a change in graph similarity. Lu simulated this scenario with a CPU anomaly that mimicked a large jump in processor utilization. Unlike the previously stable system, this system tracked a large disturbance in the graph similarity around the sample point of the CPU anomaly.

In summary, Lu et al. were able to create a system that relies on abstracting system components within a tiered application summarized by response time. This weighted graph system is able to better expose problems and anomalies in application components by uniquely utilizing security metrics, algorithms, and Euclidian graphs to be effective with a precision of up to 0.9 [12].

## V. SECURITY AUTOMATION & BEST PRACTICES

Containers are quickly taking over virtual machines. Some key reasons why container popularity is growing is because they are lightweight, portable, and ship faster [13]. However, the advantages of using containers come with vulnerabilities that are non-issues in virtual machines. In this section, the challenges associated with containers, security tools available to address and monitor these issues, and best practices to increase security are explored.

### A. Challenges

The transition from virtual machines to Docker containers comes with a host of benefits. However, the advantages of using containers also pose its own set of challenges when it comes to security. For instance, containers have a unique surface of attack compared but not unrelated to their VM counterparts. One reason is due to the fact that containers run on top of the host operation system, which they are tightly coupled to [14]. As a result, compromises within the container mean that attackers can gain access to the host OS and other local resources. Docker recent gain in popularity also means that third-party integrations and solutions are continuing to become available. As such, containers can be susceptible to malware and other vulnerabilities as mentioned in a previous section [6][13]. Finally, vulnerabilities come in their own deployment systems. For instance, one of the key benefits of Docker is the ability to deploy multiple application containers from a single service. However, even this has challenges due to the difficulty of managing several containers on a single service [13].

### B. Security Tools

A variety of tools are available for monitoring and increasing container security. One such tool is Docker Swarm. This system acts as a cluster and orchestration tool that handles automatic deployment and the detection of failed containers. Additionally, the tool includes a manager node that orchestrates and delegates processes to a worker node or nodes, which in turn runs the swarm services. The process is such that requests are distributed equally to the workers via the manager node [13]. Furthermore, Docker Swarm does have limitations when it comes to resource utilization related to hosting machines. For instance, Docker Swarm does not have the ability to directly monitor the resources on host machines, which can lead to imbalances on the host [13].

One method used for monitoring memory utilization is through the installation of scripts [13]. Scripts serve several purposes, such as recording memory usage of worker nodes and sending collected data to the manager node for analysis which can trigger the manager mode to make modifications to the configuration. However, writing scripts for each Docker Swarm can be cumbersome and inefficient due to the management nature associated with them.

Beyond writing and installing scripts, a different tool called Docker-sec, presents a more automated solution. Docker-sec works in combination with AppAmor, mentioned in the Docker Security Overview, by attaching security profiles to critical components. Docker-sec uses a two-pronged system for container security. The first part of the system is the static analysis, which is covered in detail in the section on Container Security & Metrics. In Docker-sec, the static analysis works with the security already built into containers by retrieving the initial rules that have placed restrictions on the components that containers can access. The next step requires Docker-sec to go through a training routine utilizing the dynamic monitoring mechanism. A user configures the dynamic monitoring mechanism to collect data on a container about detailed behavior over a specified time period. Once the data collection period has finished, Docker-sec analyzes the data and creates rules to manage the container profile and heighten the security. Figure 8 models the training process for a container runtime profile [15].

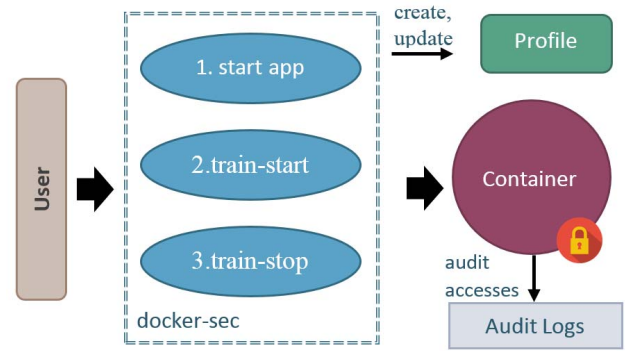

*Figure 8. Training process for a container runtime profile [13]*

*Table 9. Popular Container Security Tools [12]*

| Tool | Key Features |
| --- | --- |
| AppArmor | Container security access profiler that generates and enforces access control security. |
| AquaSec | Provides static analysis scans for vulnerabilities and security issues along with image policy enforcement and configuration management. |
| Clair | Integrates CVE analysis against container implementations |
| Docker Notary | Docker tool for the verification of image security using cryptographic signatures. |
| Docker-sec | Protects Docker containers through access policies and comparing activity with expected activity. |
| Docker Swarm | Container orchestration tool with container state management and monitoring abilities managed through the Docker Engine. |
| Sysdig | Debugging and insight tool for investigating container metrics including process, thread, and memory information. |

The enhanced container profile is implemented through the use of AppArmor in two modes. The first is enforce mode. Enforce mode establishes strict protocols, restricting the execution of any components which violate the profile rules. The other option is complain mode. Complain mode allows for executions but records the violations. The use of enforce and complain mode are not exclusive. Some profile rules can be strictly enforced while logging specific violations [15] are not exclusive. Some profile rules can be strictly enforced while logging specific violations [15]. A comparison of some of the more popular Docker security tools and their key yet distinguishing features are compared in Table 9.

*C. Security Policies*

Developers can follow several security policies to maintain and increase the level of security for Docker containers, starting with the architecture. Many of the security implications of key interest in relation to Docker are tied to its layered architecture. Docker architecture begins with the host machine. The next layer which is known as the Docker daemon or dockerd is installed on the host machine. The daemon processes and manages Docker containers, the self-contained environments that run images. The last layer is the Docker Client, which is the user interface. The Docker Client acts as a buffer between the user and the host machine. As a result, the user does not interface directly with the daemon, offering a layer of security [14].

To maintain the built-in security, developers utilize the automatically created namespaces and control groups (cgroups), allowing containers to remain isolated from one another. Next, cgroups can be used to manage container resources. This prevents containers from overusing or draining resources from other containers. Restricting access to the network and host is also important. Only the Docker Daemon should connect to the host, ensuring and enforcing the security buffer. Along with this buffer, containers should rarely have access to the root. When containers do have access to the root, privileges should be limited and scoped. In lieu of providing access or working from within root, containers can be given access to a virtual root [14].

Developers should also use third-party containers with caution. Security vulnerabilities can be coded (maliciously or unintentionally) into third-party containers, which can leave systems open to attack [16]. One often neglected item that comes with using third-party containers is in reference to the resource/package version. As mentioned in the previous section, outdated images pose significant vulnerabilities and most third-party images are outdated [6]. By using containers that are maintained, as well as keeping containers up to date, developers [17] can increase application integrity since updated containers are often equipped to handle the most recent security vulnerabilities or patches.

## VI. CONCLUSION

In summary, the containerization of applications with Docker is becoming a more popular strategy within the world of software development. With this in mind, the new landscape provides a unique level of attack surface which can be categorized by metrics key to Docker and the way in which it functions. A prime feature of containers is the images used to store and deploy applications, which utilize a GNU/Linux kernel. This results in the containers inheriting the built-in security protections or issues that come with the Linux kernel. Additional security features are available through the host operating system which has the capability of implementing Mandatory Access Controls. However, this built-in Linux and container security does not eliminate all the vulnerability concerns that come with using Docker. Specifically, this new application structure is susceptible to issues such as malicious or vulnerable images associated with third-party sharing. In fact, many of the available third-party containers have unaddressed security issues which are rated as high severity.

Additional security concerns arise from collocated containers that create access risks to other containers which can be identified by various detection scenarios and metrics. Containers are also vulnerable to various types of channel attacks such as those related to shared resources and leaks. Fortunately, many of the security challenges can be addressed or at least monitored through the use of the aforementioned utilities, metrics, and tools. Above all, Dockers unique architectural pattern provides an interesting security challenge that involves every layer of the infrastructure from image to container and generates innovative ways to collect and analyze metrics associated with the platform.

Based on the findings above, and the importance of Docker security metrics driven by the rapid adoption of containerized landscapes, several future research directions have been identified. Specifically, an advancement in the anomaly detection methods between the container and the host operating system would prove beneficial. Additionally, increased insight into adapting CVSS scoring alongside vulnerabilities specifically effecting containers could change the methods in which this system scores vulnerability severities. Finally, additional research initiatives investigating the level of Docker security education and training of enterprise architects, developers, and security experts against the rate of Docker security incidents could illuminate key methods for future incident prevention.

REFERENCES

[1] Tripwire State of Container Security Report, 3b6xlt3iddqmuq5vy2w0s5d3-wpengine.netdna-ssl.com/state-of-security/wp-content/uploads/sites/3/Tripwire-Dimensional-Research-State-of-Container-Security-Report.pdf, 2019.
[2] "Customers." Docker, www.docker.com/customers.
[3] Schreuders, Z. & McGill, Tanya & Payne, Christian. (2011). Empowering End Users to Confine Their Own Applications: The Results of a Usability Study Comparing SELinux, AppArmor, and FBAC-LSM. ACM Trans. Inf. Syst. Secur.. 14. 19. 10.1145/2019599.2019604.
[4] Schreuders, Z. & McGill, Tanya & Payne, Christian. (2011). Empowering End Users to Confine Their Own Applications: The Results of a Usability Study Comparing SELinux, AppArmor, and FBAC-LSM. ACM Trans. Inf. Syst. Secur.. 14. 19. 10.1145/2019599.2019604.
[5] Dua, Rajdeep & Raja, A & Kakadia, Dharmesh. (2014). Virtualization vs Containerization to Support PaaS. 610-614. 10.1109/IC2E.2014.41.
[6] K. Brady, S. Moon, T. Nguyen and J. Coffman, "Docker Container Security in Cloud Computing," 2020 10th Annual Computing and Communication Workshop and Conference (CCWC), Las Vegas, NV, USA, 2020, pp. 0975-0980, doi: 10.1109/CCWC47524.2020.9031195.
[7] P. Mell, K. Scarfone and S. Romanosky, The Common Vulnerability Scoring System (CVSS) and Its Applicability to Federal Agency Systems, August 2007.
[8] S. Winkel, "Security Assurance of Docker Containers: Part 1", ISSA Journal, April 2017.
[9] V. Adethyaa and T. Jernigan, "Scanning Docker Images for Vulnerabilities using Clair Amazon ECS ECR and AWS CodePipeline", AWS Compute Blog, November 2018.

[10] A. Zerouali, T. Mens, G. Robles and J. M. Gonzalez-Barahona, "On the Relation between Outdated Docker Containers, Severity Vulnerabilities, and Bugs," 2019 IEEE 26th International Conference on Software Analysis, Evolution and Reengineering (SANER), Hangzhou, China, 2019, pp. 491-501, doi: 10.1109/SANER.2019.8668013.

[11] Y. Luo, W. Luo, X. Sun, Q. Shen, A. Ruan and Z. Wu, "Whispers between the Containers: High-Capacity Covert Channel Attacks in Docker," 2016 IEEE Trustcom/BigDataSE/ISPA, Tianjin, 2016, pp. 630-637, doi: 10.1109/TrustCom.2016.0119.

[12] C. Lu, K. Ye, W. Chen and C. Xu, "ADGS: Anomaly Detection and Localization Based on Graph Similarity in Container-Based Clouds," 2019 IEEE 25th International Conference on Parallel and Distributed Systems (ICPADS), Tianjin, China, 2019, pp. 53-60, doi: 10.1109/ICPADS47876.2019.00016.

[13] M. R. M. Bella, M. Data and W. Yahya, "Web Server Load Balancing Based On Memory Utilization Using Docker Swarm," 2018 International Conference on Sustainable Information Engineering and Technology (SIET), Malang, Indonesia, 2018, pp. 220-223, doi: 10.1109/SIET.2018.8693212.

[14] A. R. Manu, J. K. Patel, S. Akhtar, V. K. Agrawal and K. N. B. Subramanya Murthy, "A study, analysis and deep dive on cloud PAAS security in terms of Docker container security," 2016 International Conference on Circuit, Power and Computing Technologies (ICCPCT), Nagercoil, 2016, pp. 1-13, doi: 10.1109/ICCPCT.2016.7530284.

[15] F. Loukidis-Andreou, I. Giannakopoulos, K. Doka and N. Koziris, "Docker-Sec: A Fully Automated Container Security Enhancement Mechanism," 2018 IEEE 38th International Conference on Distributed Computing Systems (ICDCS), Vienna, 2018, pp. 1561-1564, doi: 10.1109/ICDCS.2018.00169.

[16] H. HaddadPajouh, A. Dehghantanha, R. M. Parizi, M. Aledhari, H. Karimipour, "A survey on internet of things security: Requirements, challenges, and solutions", Internet of Things, 2019, https://doi.org/10.1016/j.iot.2019.100129.

[17] R. M. Parizi, "On the gamification of human-centric traceability tasks in software testing and coding," 2016 IEEE 14th International Conference on Software Engineering Research, Management and Applications (SERA), Towson, MD, 2016, pp. 193-200, doi: 10.1109/SERA.2016.7516146.

[18] R. M. Parizi, K. Qian, H. Shahriar, F. Wu and L. Tao, "Benchmark Requirements for Assessing Software Security Vulnerability Testing Tools," 2018 IEEE 42nd Annual Computer Software and Applications Conference (COMPSAC), Tokyo, 2018, pp. 825-826, doi: 10.1109/COMPSAC.2018.00139.